\documentclass[aps,twocolumn,pra,superscriptaddress,showpacs,tightenlines]{revtex4}
\usepackage{epsfig,graphicx,times}
\usepackage{amstext}
\usepackage{amsmath}
\usepackage{amssymb}
\usepackage{graphicx}
\usepackage{latexsym}
\usepackage{bm}
\usepackage[colorlinks,citecolor=blue, linkcolor=blue,hyperindex,CJKbookmarks,dvipdfm]{hyperref}

\begin{document}

\title{Optical nonreciprocity and optomechanical circulator in three-mode
optomechanical systems}
\author{Xun-Wei Xu}
\affiliation{Beijing Computational Science Research Center, Beijing 100084, China}
\author{Yong Li}
\email{liyong@csrc.ac.cn}
\affiliation{Beijing Computational Science Research Center, Beijing 100084, China}
\affiliation{Synergetic Innovation Center of Quantum Information and Quantum Physics,
University of Science and Technology of China, Hefei, Anhui 230026, China}
\date{\today }

\begin{abstract}
We demonstrate the possibility of optical nonreciprocal response in a
three-mode optomechanical system where one mechanical mode is
optomechanically coupled to two linearly coupled optical modes
simultaneously. The optical nonreciprocal behavior is induced by the phase
difference between the two optomechanical coupling rates which breaks the
time-reversal symmetry of the three-mode optomechanical system. Moreover,
the three-mode optomechanical system can also be used as a three-port
circulator for two optical and one mechanical modes, which we refer to as optomechanical circulator.
\end{abstract}

\pacs{42.50.Wk, 42.50.Ex, 07.10.Cm, 11.30.Er}
\maketitle

%42.50.Wk Mechanical effects of light on material media, microstructures and particles (see also 87.80.Cc Optical trapping in biology and medicine)
%42.50.Ex Optical implementations of quantum information processing and transfer
%07.10.Cm Micromechanical devices and systems [for micro- and nano-electromechanical systems (MEMS/NEMS)
%11.30.Er Charge conjugation, parity, time reversal, and other discrete symmetries

\section{Introduction}

The fundamental role of nonreciprocal transmission in information processing
has been demonstrated fully by the important application of electrical
diodes in electronic information technology with semiconductor p-n
junctions. However, optical nonreciprocity is constrained by the Lorentz
reciprocal theorem due to the time-reversal symmetry in linear and nonmagnetic media~\cite{PottonRPP04}. Traditionally, optical nonreciprocity is based on
magneto-optical crystals~\cite{FujitaAPL00} by breaking the time-reversal
symmetry with the Faraday rotation effect, or optical nonlinear systems~\cite{GalloAPL01} by circumventing the symmetrical constraint. Recently, a number of alternative schemes based on diverse mechanisms have been proposed, such as spatial-symmetry-breaking structures~\cite{BiancalanaJAP08}, indirect
interband photonic transitions~\cite{YuNP09}, opto-acoustic effects~\cite{WangOE10}, parity-time symmetric structures~\cite{EuterNP10}, and moving systems~\cite{WangPRL13}. Moreover, for the potential applications in
photonic quantum information processing, the realization of nonreciprocal
photonic devices with the ability to be integrated on a chip and operating
on a single-photon level~\cite{ShenPRL11} are desirable features in future.

With rapidly growing interest as a new class of microscale integratable devices, optomechanical systems have shown enormous potential for the application in quantum information processing~\cite{AspelmeyerARX13}.
%It has already been shown that optomechanical systems can be used to induce nonreciprocal effects for light by the intrinsic nonlinearity [11] and external driving fields [12].
It has already been shown that optomechanical systems can be used to induce
nonreciprocal effects for light~\cite{ManipatruniPRL09,HafeziOE12}. At the
beginning, the optical nonreciprocal effect is based on the momentum
difference between forward and backward moving light beams in the optomechanical
system consisting of an inline Fabry-P\'{e}rot cavity with one movable
mirror and one fixed mirror~\cite{ManipatruniPRL09}. Subsequently a new
approach for nonreciprocal optomechanical device was proposed by using
strong optomechanical interaction in microring resonators~\cite{HafeziOE12}.
The nonreciprocal response is obtained for the optomechanical coupling is
enhanced in one direction and suppressed in the other one by optically
pumping the ring resonator. In principle, the scheme shown in Ref.~\cite{HafeziOE12} can be applied on a single-photon level, in spite of the limitation induced by the up-conversion of thermal phonons.

In this paper, we propose a scheme for optical nonreciprocity in a
three-mode optomechanical system, where two optical modes are linearly
coupled to each other and one mechanical mode is optomechanically coupled to
the two optical modes simultaneously. The two effective optomechanical
couplings are both enhanced by pumping the two optical modes with different
external driving fields, respectively. And most crucially, there is a phase
difference between the two effective optomechanical couplings, which cannot
be absorbed into local redefinitions of the operators. Nonreciprocal
response of the three-mode optomechanical system is induced by this phase
difference which can be associated with an effective magnetic field for the
three modes~\cite{KochPRA10,HabrakenNJP12}. This mechanism has been used in the circuit-QED architecture~\cite{KochPRA10} and phonon device~\cite{HabrakenNJP12} for breaking time-reversal symmetry, and photon or phonon circulator behavior~\cite{KochPRA10,HabrakenNJP12} was predicted accordingly.
Thus, the present three-mode optomechanical system can also be used as a three-port circulator %and it is worth mentioning that the three-port circulator in the system here is
formed by two optical and one mechanical modes, which we refer to as optomechanical circulator. This new type of circulators may serve as suitable interfaces for the hybrid network comprised of optical (or microwave) and mechanical systems.

This paper is organized as follows: In Sec.~II, the Hamiltonian of a
three-mode optomechanical system is introduced and the spectra of the
output fields are obtained formally. The optical nonreciprocal response is
shown in Sec.~III and the optomechanical circulator behavior is discussed in
Sec.~IV. Finally, we draw our conclusions in Sec.~V.

\section{Model}

\begin{figure}[tbp]
\includegraphics[bb=75 285 524 496, width=8 cm, clip]{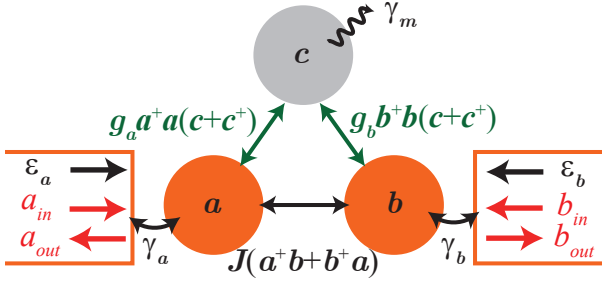}
\caption{(Color online) Schematic diagram of an optomechanical system
consisting of two optical modes ($a$ and $b$) and one mechanical mode ($c$).
The optical modes and the mechanical mode are coupled via radiation pressure, respectively; meanwhile, the two optical modes are linearly coupled to each other.}
\label{fig1}
\end{figure}

We consider a three-mode optomechanical system~\cite{LudwigPRL12} consisting
of two optical modes ($a$ and $b$, frequencies $\omega _{a}$ and $\omega _{b}
$) and one mechanical mode ($c$, frequency $\omega _{m}$) as shown in Fig.~\ref{fig1}. The optomechanical coupling rates between the optical modes and
the mechanical mode are denoted by $g_{a}$ and $g_{b}$; two optical modes
are linearly coupled mutually at rate $J$ and driven
by external laser sources with frequencies $\omega _{a,d}=\omega _{b,d}=\omega _{d}$ at rates $\varepsilon _{a}$ and $\varepsilon _{b}$ respectively. The Hamiltonian of the system in the rotating frame at the frequency of the driving fields $\omega _{d}$ is
\begin{eqnarray}
H &=&\hbar \Delta _{a}a^{\dag }a+\hbar \Delta _{b}b^{\dag }b+\hbar \omega
_{m}c^{\dag }c+\hbar J\left( a^{\dag }b+ab^{\dag }\right)   \notag \\
&&+\hbar g_{a}a^{\dag }a\left( c+c^{\dag }\right) +\hbar g_{b}b^{\dag
}b\left( c+c^{\dag }\right)   \notag \\
&&+i\hbar \left( \varepsilon _{a}a^{\dag }e^{i\phi _{a}}+\varepsilon
_{b}b^{\dag }e^{i\phi _{b}}-\mathrm{H.c.}\right) ,  \label{Eq1}
\end{eqnarray}%
where $\Delta _{a}=\omega _{a}-\omega _{d}$ and $\Delta _{b}=\omega_{b}-\omega _{d}$ are the detunings between the optical modes and the driving fields. Without loss of generality, we assume that $J$, $\varepsilon _{a}$ and $\varepsilon _{b}$ are real and $\phi _{a} $ ($\phi _{b}$) is the phase of laser field coupling to the optical mode $a$ ($b$). This kind of Hamiltonian can be realized in the optomechanical system with a membrane in a Fabry-P\'{e}rot cavity~\cite{ThompsonNat08}, microtoroid optomechanical cavities~\cite{LinPRL09}, optomechanical crystals~\cite{EichenfieldNat09} and electromechanical devices~\cite{TeufelNat11}.

Substituting the Hamiltonian~(\ref{Eq1}) into the Heisenberg equation and
taking into account the damping and corresponding noise terms, we get the
quantum Langevin equations (QLEs) for the operators of the optical and
mechanical modes
\begin{eqnarray}
\frac{d}{dt}a &=&\left\{ -\frac{\gamma _{a}}{2}-i\left[ \Delta
_{a}+g_{a}\left( c+c^{\dag }\right) \right] \right\} a-iJb  \notag
\label{Eq2} \\
&&+\varepsilon _{a}e^{i\phi _{a}}+\sqrt{\gamma _{a}}a_{\mathrm{in}},
\end{eqnarray}%
\begin{eqnarray}
\frac{d}{dt}b &=&\left\{ -\frac{\gamma _{b}}{2}-i\left[ \Delta
_{b}+g_{b}\left( c+c^{\dag }\right) \right] \right\} b-iJa  \notag
\label{Eq3} \\
&&+\varepsilon _{b}e^{i\phi _{b}}+\sqrt{\gamma _{b}}b_{\mathrm{in}},
\end{eqnarray}%
\begin{equation}
\frac{d}{dt}c=\left( -\frac{\gamma _{m}}{2}-i\omega _{m}\right) c-i\left(
g_{a}a^{\dag }a+g_{b}b^{\dag }b\right) +\sqrt{\gamma _{m}}c_{\mathrm{in}}.
\label{Eq4}
\end{equation}%
Here, $\gamma _{a}$ ($\gamma _{b}$) is the damping rate of the optical mode $a$ ($b$) and $\gamma _{m}$ is the mechanical damping rate. $a_{\mathrm{in}}$%
, $b_{\mathrm{in}}$ and $c_{\mathrm{in}}$ are the input quantum fields with zero mean values, and the spectra of the input quantum fields, $s_{v,\mathrm{in}}\left( \omega \right)$, are defined via $\left\langle \widetilde{v_{\mathrm{in}}^{\dag }}\left( \omega ^{\prime }\right) \widetilde{v_{\mathrm{in}}}\left(\omega \right) \right\rangle =s_{v,\mathrm{in}}\left( \omega \right) \delta
\left( \omega +\omega ^{\prime }\right) $ and $\left\langle \widetilde{v_{\mathrm{in} }}\left( \omega ^{\prime }\right) \widetilde{v_{\mathrm{in}}^{\dag }}\left(
\omega \right) \right\rangle =\left[ 1+s_{v,\mathrm{in}}\left( \omega \right)\right] \delta \left( \omega +\omega ^{\prime }\right) $, where the term ``1'' results from the effect of vacuum noise and $\widetilde{v_{\mathrm{in}
}^{\dag }}$ ($\widetilde{v_{\mathrm{in}}}$) is the Fourier transform of $
v_{\mathrm{in}}^{\dag }$ ($v_{\mathrm{in}}$) for $v=a,b,c$.

The mean values of the operators in the steady state can be obtained from
the nonlinear QLEs (\ref{Eq2})-(\ref{Eq4}) by using factorization assumption like $\left\langle ca\right\rangle =\left\langle c\right\rangle \left\langle
a\right\rangle $, and then
\begin{eqnarray}
\left\langle a\right\rangle  &=&\alpha =\frac{\left( \frac{\gamma _{b}}{2}%
+i\Delta _{b}^{\prime }\right) \varepsilon _{a}e^{i\phi _{a}}-iJ\varepsilon
_{b}e^{i\phi _{b}}}{\left( \frac{\gamma _{a}}{2}+i\Delta _{a}^{\prime
}\right) \left( \frac{\gamma _{b}}{2}+i\Delta _{b}^{\prime }\right) +J^{2}},
\label{Eq5} \\
\left\langle b\right\rangle  &=&\beta =\frac{\left( \frac{\gamma _{a}}{2}%
+i\Delta _{a}^{\prime }\right) \varepsilon _{b}e^{i\phi _{b}}-iJ\varepsilon
_{a}e^{i\phi _{a}}}{\left( \frac{\gamma _{a}}{2}+i\Delta _{a}^{\prime
}\right) \left( \frac{\gamma _{b}}{2}+i\Delta _{b}^{\prime }\right) +J^{2}},
\label{Eq6} \\
\left\langle c\right\rangle  &=&\xi =\frac{-i\left( g_{a}\left\vert \alpha
\right\vert ^{2}+g_{b}\left\vert \beta \right\vert ^{2}\right) }{\left(
\frac{\gamma _{m}}{2}+i\omega _{m}\right) },  \label{Eq7}
\end{eqnarray}%
where $\Delta _{a}^{\prime }=\Delta _{a}+g_{a}\left( \xi +\xi ^{\ast
}\right) $ and $\Delta _{b}^{\prime }=\Delta _{b}+g_{b}\left( \xi +\xi
^{\ast }\right) $ are the effective detuning including the frequency shifts
caused by the optomechanical interaction.

To solve the nonlinear QLEs (\ref{Eq2})-(\ref{Eq4}), we linearize the
equations in the strong driving condition (i.e., $\varepsilon _{a}\gg \gamma
_{a}$, $\varepsilon _{b}\gg \gamma _{b}$), then the operators are rewritten
as the sum of the mean values and the small quantum fluctuation terms i.e., $a=\alpha +\delta a$, $b=\beta +\delta b$, $c=\xi +\delta c$, where $\delta
a\ll |\alpha |$ and $\delta b\ll |\beta |$. Substituting them into the nonlinear
QLEs (\ref{Eq2})-(\ref{Eq4}) and keeping only the first-order terms in the
small quantum fluctuation terms $\delta a$, $\delta b$, and $\delta c$, we
obtain the linearized QLEs
\begin{eqnarray}
\frac{d}{dt}\delta a &=&\left( -\frac{\gamma _{a}}{2}-i\Delta _{a}^{\prime
}\right) \delta a-iG_{a}\left( \delta c+\delta c^{\dag }\right)  \notag \\
&&-iJ\delta b+\sqrt{\gamma _{a}}a_{\mathrm{in}},  \label{Eq8} \\
\frac{d}{dt}\delta b &=&\left( -\frac{\gamma _{b}}{2}-i\Delta _{b}^{\prime
}\right) \delta b-iG_{b}\left( \delta c+\delta c^{\dag }\right)  \notag \\
&&-iJ\delta a+\sqrt{\gamma _{b}}b_{\mathrm{in}},  \label{Eq9} \\
\frac{d}{dt}\delta c &=&\left( -\frac{\gamma _{m}}{2}-i\omega _{m}\right)
\delta c-i\left( G_{a}\delta a^{\dag }+G_{a}^{\ast }\delta a\right)  \notag
\\
&&-i\left( G_{b}\delta b^{\dag }+G_{b}^{\ast }\delta b\right) +\sqrt{\gamma
_{m}}c_{\mathrm{in}},  \label{Eq10}
\end{eqnarray}%
where $G_{a}=g_{a}\alpha= |G_{a}|e^{i\theta _{a}}$ and $G_{b}=g_{b}\beta
=|G_{b}|e^{i\theta _{b}}$ are the effective optomechanical coupling rates
with phase difference $\theta\equiv\theta _{b}-\theta _{a}$.

For convenience, the linearized QLEs (\ref{Eq8})-(\ref{Eq10}) can be
concisely expressed as
\begin{equation}
\frac{d}{dt}V=-MV+\Gamma V_{\mathrm{in}},  \label{Eq11}
\end{equation}%
where the fluctuation vector $V=\left( \delta a,\delta b,\delta c,\delta a^{\dag
},\delta b^{\dag },\delta c^{\dag }\right) ^{T}$, the input field vector $%
V_{in}=\left( a_{\mathrm{in}},b_{\mathrm{in}},c_{\mathrm{in}},a_{\mathrm{in}%
}^{\dag },b_{\mathrm{in}}^{\dag },c_{\mathrm{in}}^{\dag }\right) ^{T}$, $%
\Gamma =\mathrm{diag}\left( \sqrt{\gamma _{a}},\sqrt{\gamma _{b}},\sqrt{%
\gamma _{m}},\sqrt{\gamma _{a}},\sqrt{\gamma _{b}},\sqrt{\gamma _{m}}\right)
$ denotes the damping matrix and $M$ is the coefficient matrix
\begin{widetext}
\begin{equation}
M=\left(
\begin{array}{cccccc}
\frac{\gamma _{a}}{2}+i\Delta _{a}^{\prime } & iJ & iG_{a} & 0 & 0 & iG_{a}
\\
iJ & \frac{\gamma _{b}}{2}+i\Delta _{b}^{\prime } & iG_{b} & 0 & 0 & iG_{b}
\\
iG_{a}^{\ast } & iG_{b}^{\ast } & \frac{\gamma _{m}}{2}+i\omega _{m} & iG_{a}
& iG_{b} & 0 \\
0 & 0 & -iG_{a}^{\ast } & \frac{\gamma _{a}}{2}-i\Delta _{a}^{\prime } & -iJ
& -iG_{a}^{\ast } \\
0 & 0 & -iG_{b}^{\ast } & -iJ & \frac{\gamma _{b}}{2}-i\Delta _{b}^{\prime }
& -iG_{b}^{\ast } \\
-iG_{a}^{\ast } & -iG_{b}^{\ast } & 0 & -iG_{a} & -iG_{b} & \frac{\gamma _{m}
}{2}-i\omega _{m}
\end{array}
\right) .  \label{Eq12}
\end{equation}
\end{widetext}%
Due to the stability condition, the real parts of all the eigenvalues of matrix $M$ have to be positive. By introducing the Fourier transform of the operators
\begin{eqnarray}
\widetilde{o}\left( \omega \right) &=&\frac{1}{\sqrt{2\pi }}\int_{-\infty
}^{+\infty }o\left( t\right) e^{i\omega t}dt,  \label{Eq13} \\
\widetilde{o^{\dag }}\left( \omega \right) &=&\frac{1}{\sqrt{2\pi }}%
\int_{-\infty }^{+\infty }o^{\dag }\left( t\right) e^{i\omega t}dt,
\label{Eq14}
\end{eqnarray}%
(for any operator $o$) and using the properties of Fourier transformation, the solution to the linearized QLEs (\ref{Eq11}) in the frequency domain is
\begin{equation}
\widetilde{V}\left( \omega \right) =\left( M-i\omega I\right) ^{-1}\Gamma
\widetilde{V}_{\mathrm{in}}\left( \omega \right) ,  \label{Eq15}
\end{equation}%
where $I$ denotes the identity matrix.

As a consequence of boundary conditions, the relation among the input,
internal, and output fields is given as the fol1owing~\cite{GardinerPRA85}
\begin{equation}
v_{\mathrm{out}}+v_{\mathrm{in}}=\sqrt{\gamma _{v}}\delta v  \label{Eq16}
\end{equation}%
for $v=a,b,c$, and $\gamma _{c}\equiv \gamma _{m}$. Then the output field vector
in the frequency domain is
\begin{equation}
\widetilde{V}_{\mathrm{out}}\left( \omega \right) =U\left( \omega \right)
\widetilde{V}_{\mathrm{in}}\left( \omega \right) ,  \label{Eq17}
\end{equation}%
where the output field vector $\widetilde{V}_{\mathrm{out}}\left( \omega \right)$ is the Fourier
transform of $V_{\mathrm{out}}=\left( a_{\mathrm{out}},b_{\mathrm{out}},c_{%
\mathrm{out}},a_{\mathrm{out}}^{\dag },b_{\mathrm{out}}^{\dag },c_{\mathrm{%
out}}^{\dag }\right) ^{T}$ and
\begin{equation}
U\left( \omega \right) =\Gamma \left( M-i\omega I\right) ^{-1}\Gamma -I.
\label{Eq18}
\end{equation}%
The spectrum of the output fields is defined by
\begin{equation}
s_{v,\mathrm{out}}\left( \omega \right) =\int d\omega ^{\prime }\left\langle
\widetilde{v_{\mathrm{out}}^{\dag }}\left( \omega^{\prime } \right) \widetilde{v_{%
\mathrm{out}}}\left( \omega\right) \right\rangle .  \label{Eq19}
\end{equation}%
By substituting the expression of $\widetilde{V}_{\mathrm{out}}\left( \omega
\right) $ [Eq.~(\ref{Eq17})] into Eq.~(\ref{Eq19}), one can obtain~\cite%
{AgarwalPRA12}
\begin{equation}
S_{\mathrm{out}}\left( \omega \right) =T\left( \omega \right) S_{\mathrm{in}%
}\left( \omega \right) +S_{\mathrm{vac}}\left( \omega \right) .  \label{Eq20}
\end{equation}%
Here $S_{\mathrm{in}}\left( \omega \right) =\left( s_{a,\mathrm{in}}\left(
\omega \right) ,s_{b,\mathrm{in}}\left( \omega \right) ,s_{c,\mathrm{in}%
}\left( \omega \right) \right) ^{T}$, $S_{\mathrm{out}}\left( \omega \right)
=\left( s_{a,\mathrm{out}}\left( \omega \right) ,s_{b,\mathrm{out}}\left(
\omega \right) ,s_{c,\mathrm{out}}\left( \omega \right) \right) ^{T}$, $S_{%
\mathrm{vac}}\left( \omega \right) =\left( s_{a,\mathrm{vac}}\left( \omega
\right) ,s_{b,\mathrm{vac}}\left( \omega \right) ,s_{c,\mathrm{vac}}\left(
\omega \right) \right) ^{T}$, and
\begin{equation}
T\left( \omega \right) =\left(
\begin{array}{ccc}
T_{aa}\left( \omega \right)  & T_{ab}\left( \omega \right)  & T_{ac}\left(
\omega \right)  \\
T_{ba}\left( \omega \right)  & T_{bb}\left( \omega \right)  & T_{bc}\left(
\omega \right)  \\
T_{ca}\left( \omega \right)  & T_{cb}\left( \omega \right)  & T_{cc}\left(
\omega \right)
\end{array}%
\right) ,  \label{Eq21}
\end{equation}%
where the element $T_{ij}\left( \omega \right) $ ($i,j=a,b,c$) denotes the
scattering probability that is corresponding to the output field of $i$ mode arising from the presence of a single photon (or single phonon) in the
input field of $j$ mode. The scattering probabilities are given as
\begin{eqnarray}
T_{aa}\left( \omega \right)  &=&\left\vert U_{11}\left( \omega \right)
\right\vert ^{2}+\left\vert U_{14}\left( \omega \right) \right\vert ^{2},
\label{Eq22} \\
T_{ab}\left( \omega \right)  &=&\left\vert U_{12}\left( \omega \right)
\right\vert ^{2}+\left\vert U_{15}\left( \omega \right) \right\vert ^{2},
\label{Eq23} \\
T_{ac}\left( \omega \right)  &=&\left\vert U_{13}\left( \omega \right)
\right\vert ^{2}+\left\vert U_{16}\left( \omega \right) \right\vert ^{2},
\label{Eq24} \\
T_{ba}\left( \omega \right)  &=&\left\vert U_{21}\left( \omega \right)
\right\vert ^{2}+\left\vert U_{24}\left( \omega \right) \right\vert ^{2},
\label{Eq25} \\
T_{bb}\left( \omega \right)  &=&\left\vert U_{22}\left( \omega \right)
\right\vert ^{2}+\left\vert U_{25}\left( \omega \right) \right\vert ^{2},
\label{Eq26} \\
T_{bc}\left( \omega \right)  &=&\left\vert U_{23}\left( \omega \right)
\right\vert ^{2}+\left\vert U_{26}\left( \omega \right) \right\vert ^{2},
\label{Eq27} \\
T_{ca}\left( \omega \right)  &=&\left\vert U_{31}\left( \omega \right)
\right\vert ^{2}+\left\vert U_{34}\left( \omega \right) \right\vert ^{2},
\label{Eq28} \\
T_{cb}\left( \omega \right)  &=&\left\vert U_{32}\left( \omega \right)
\right\vert ^{2}+\left\vert U_{35}\left( \omega \right) \right\vert ^{2},
\label{Eq29} \\
T_{cc}\left( \omega \right)  &=&\left\vert U_{33}\left( \omega \right)
\right\vert ^{2}+\left\vert U_{36}\left( \omega \right) \right\vert ^{2},
\label{Eq30}
\end{eqnarray}
where $U_{ij}\left( \omega \right)$ (for $i,j=1,\cdots,6$) represents the element at the $i$th row and $j$th column of the matrix $U\left( \omega \right)$ given by Eq.~(\ref{Eq18}).
$s_{v,\mathrm{vac}}$ ($v=a,b,c$) is the output spectrum contributing from the input
vacuum field,
\begin{eqnarray}
s_{a,\mathrm{vac}}\left( \omega \right)  &=&\left\vert U_{14}\left( \omega
\right) \right\vert ^{2}+\left\vert U_{15}\left( \omega \right) \right\vert
^{2}+\left\vert U_{16}\left( \omega \right) \right\vert ^{2},  \label{Eq31}
\\
s_{b,\mathrm{vac}}\left( \omega \right)  &=&\left\vert U_{24}\left( \omega
\right) \right\vert ^{2}+\left\vert U_{25}\left( \omega \right) \right\vert
^{2}+\left\vert U_{26}\left( \omega \right) \right\vert ^{2},  \label{Eq32}
\\
s_{c,\mathrm{vac}}\left( \omega \right)  &=&\left\vert U_{34}\left( \omega
\right) \right\vert ^{2}+\left\vert U_{35}\left( \omega \right) \right\vert
^{2}+\left\vert U_{36}\left( \omega \right) \right\vert ^{2}.  \label{Eq33}
\end{eqnarray}

\section{Optical nonreciprocity}

\begin{widetext}
\begin{figure*}[tbp]
\includegraphics[bb=0 307 591 553, width=16.5 cm, clip]{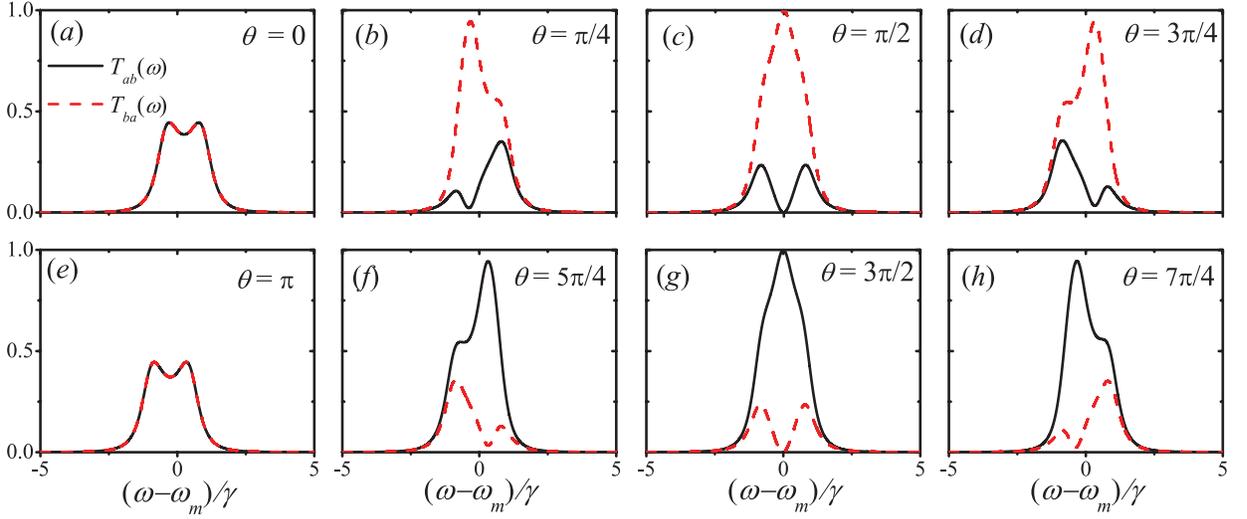}
\caption{(Color online) Scattering probabilities $T_{ab}\left(
\omega \right)$ (black solid lines) and $T_{ba}\left( \omega \right)$
(red dash lines) as functions of the frequency of the incoming signal $\omega$ for different phase difference: (a) $\theta =0$; (b) $
\theta =\pi/4$; (c) $\theta =\pi/2$; (d) $
\theta =3\pi/4$; (e) $\theta =\pi$; (f) $
\theta =5\pi/4$; (g) $\theta =3\pi/2$; (h) $\theta
=7\pi/4$. The other parameters are $\Delta _{a}^{\prime }=\Delta
_{b}^{\prime }=\omega _{m}=10\gamma$, $J=G_{a}=G_{b}e^{-i\theta }=
\gamma _{a}/2=\gamma _{b}/2=\gamma _{m}/2= \gamma /2$.}
\label{fig2}
\end{figure*}
\end{widetext}

In this and next sections, we numerically evaluate the scattering probabilities to show the possibility of optical nonreciprocal response and optomechanical circulator behavior in the three-mode optomechanical system. The optimal parameters for the observation of optical nonreciprocal response are obtained according to the numerical results. The physical origin for the optical nonreciprocal response and optomechanical circulator behavior will be discussed in the next section.

Scattering probabilities $T_{ab}\left( \omega \right) $ and $T_{ba}\left(
\omega \right) $ as functions of the frequency of the incoming signal $%
\omega $ for different phase difference are shown in Fig.~\ref{fig2}, where
the parameters are $\Delta _{a}^{\prime }=\Delta _{b}^{\prime }=\omega
_{m}=10\gamma $, $J=G_{a}=G_{b}e^{-i\theta }=\gamma _{a}/2=\gamma
_{b}/2=\gamma _{m}/2=\gamma /2$. The photon transmission satisfies the
Lorentz reciprocal theorem [e.g. $T_{ab}\left( \omega \right) =T_{ba}\left(
\omega \right) $] on the condition that $\theta =0$ or $\pi $. In the regime
$0<\theta <\pi $, we have $T_{ab}\left( \omega \right) < T_{ba}\left( \omega
\right) $; in the regime $\pi <\theta <2\pi $, we have $T_{ab}\left( \omega
\right) > T_{ba}\left( \omega \right) $. The optimal optical nonreciprocal
response is obtained as $\theta =\pi /2$ [$T_{ab}\left( \omega \right)
\approx 0$ and $T_{ba}\left( \omega \right) \approx 1$ at $\omega =\omega
_{m}$] and $\theta =3\pi /2$ [$T_{ab}\left( \omega \right) \approx 1$ and $%
T_{ba}\left( \omega \right) \approx 0$ at $\omega =\omega _{m}$]. The
condition of $G_{a}=G_{b}e^{-i\theta }=\gamma /2$ with $g_{a}=g_{b}=g$ can be obtained approximately by setting the amplitudes and the phases of the coupling laser fields as
\begin{equation}
\varepsilon _{a}=\varepsilon _{b}\approx \frac{\gamma \omega _{m}}{2g},
\label{Eq34}
\end{equation}%
\begin{equation}
\phi _{a}=\phi _{b}-\theta \approx \frac{\pi }{2}.  \label{Eq35}
\end{equation}

\begin{figure}[tbp]
\includegraphics[bb=62 243 517 608, width=8.5 cm, clip]{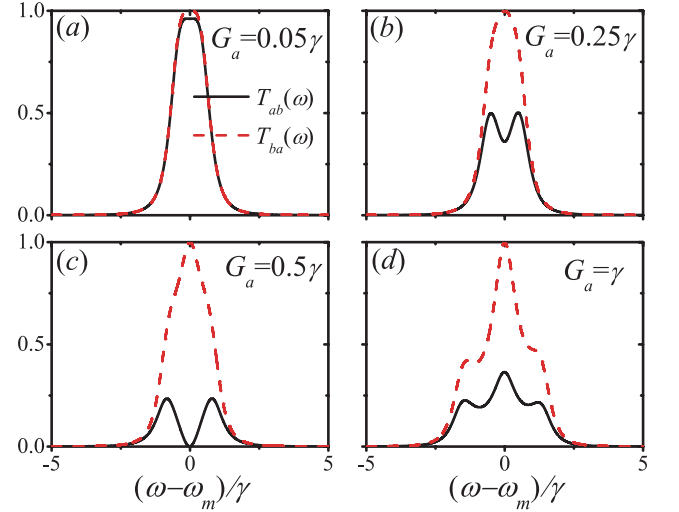}
\caption{(Color online) Scattering probabilities $T_{ab}\left( \protect%
\omega \right)$ (black solid line) and $T_{ba}\left( \protect\omega \right)$
(red dash line) as functions of the frequency of the incoming signal $%
\protect\omega$ for different effective optomechanical coupling rates: (a) $%
G _{a} =0.05 \protect\gamma $; (b) $G _{a} =0.25 \protect\gamma $; (c) $G
_{a} =0.5 \protect\gamma $; (d) $G _{a} = \protect\gamma $. The other
parameters are $\Delta _{a}^{\prime }=\Delta _{b}^{\prime }=\protect\omega %
_{m}=10\protect\gamma$, $J=\protect\gamma _{a}/2=\protect\gamma _{b}/2=
\protect\gamma _{m}/2 =\protect\gamma /2$. Here we fix $G_{b}=iG _{a}$ which corresponds to the case of $\protect\theta=\protect\pi/2$.}
\label{fig3}
\end{figure}

In Fig.~\ref{fig3}, the scattering probabilities $T_{ab}\left( \omega
\right) $ and $T_{ba}\left( \omega \right)$ are shown as functions of the
frequency of the incoming signal $\omega$ for different effective
optomechanical coupling rates $G _{a}$ with the parameters: $\Delta _{a}^{\prime
}=\Delta _{b}^{\prime }=\omega _{m}=10\gamma$, $J=\gamma _{a}/2=\gamma
_{b}/2=\gamma _{m}/2 =\gamma /2$, $G_{b}=iG _{a}$. It is shown that as the
effective optomechanical coupling is weak ($\{|G _{a}|,|G _{b}|\} \ll \gamma$), the scattering probability from $b$ mode to $a$ mode is almost the same
as the one from $a$ mode to $b$ mode, e.g., $T_{ab}\left( \omega
\right)\approx T_{ba}\left( \omega \right)$. With the enhancement of the
effective optomechanical coupling rates, the optical nonreciprocal response
becomes obvious and gets to the optimal effect at about $G _{a} =0.5 \gamma $.

\begin{figure}[tbp]
\includegraphics[bb=70 245 511 606, width=8.5 cm, clip]{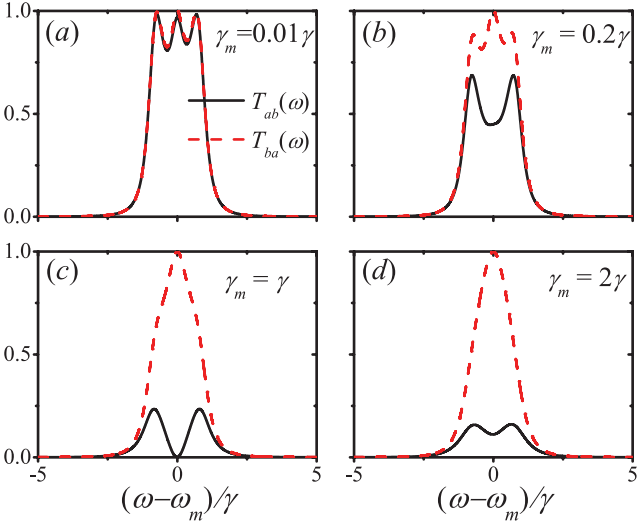}
\caption{(Color online) Scattering probabilities $T_{ab}\left( \protect%
\omega \right)$ (black solid lines) and $T_{ba}\left( \protect\omega \right)$
(red dash lines) as functions of the frequency of the incoming signal $%
\protect\omega$ for different mechanical damping rates: (a) $\protect\gamma %
_{m} =0.01 \protect\gamma$; (b) $\protect\gamma _{m} =0.2 \protect\gamma$;
(c) $\protect\gamma _{m} = \protect\gamma$; (d) $\protect\gamma _{m} =2
\protect\gamma$. The other parameters are $\Delta _{a}^{\prime }=\Delta
_{b}^{\prime }=\protect\omega _{m}=10\protect\gamma$, $J=G_{a}=-iG_{b}=%
\protect\gamma _{a}/2=\protect\gamma _{b}/2=\protect\gamma /2$.}
\label{fig4}
\end{figure}

In Fig.~\ref{fig4}, we plot the scattering probabilities $T_{ab}\left(
\omega \right)$ and $T_{ba}\left( \omega \right)$ for different mechanical
damping rates $\gamma _{m}$ with the parameters: $\Delta _{a}^{\prime }=\Delta
_{b}^{\prime }=\omega _{m}=10\gamma$, $J=G_{a}=-iG_{b}=\gamma _{a}/2=\gamma
_{b}/2=\gamma /2$. It is shown that as the mechanical damping rates $\gamma
_{m}$ is much smaller than the optical damping rate ($\gamma _{m} \ll \gamma$), the photon scattering probabilities are almost the same for the two
directions, e.g. $T_{ab}\left( \omega \right)\approx T_{ba}\left( \omega
\right)$. With the increase of the mechanical damping rate, the optical
nonreciprocal response becomes obvious and achieves the optimal effect for $\gamma _{m} \approx \gamma$.

\section{Optomechanical circulator}

As done in most of the studies on optomechanical systems, the signal
input and/or output from the mechanical mode is not considered in last section.
With the development of phonon-based system, phonon is another useful media
for quantum information processing~\cite{HabrakenNJP12}. In this section, we
assume that the mechanical mode is coupled to a continuous mode of phonon
waveguide and the phonons can be input and output through the phonon waveguide.
The scattering of both photons and phonons in the three-mode optomechanical
system is considered in the following.

\begin{figure}[tbp]
\includegraphics[bb= 18 101 560 702, width=8.5 cm, clip]{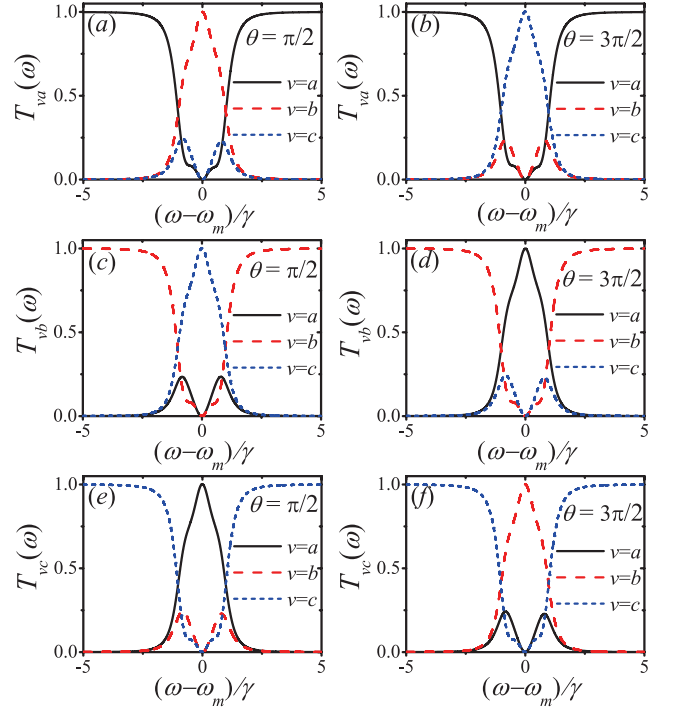}
\caption{(Color online) Scattering probabilities $T_{va}\left( \protect%
\omega \right)$ [(a) and (b)], $T_{vb}\left( \protect\omega \right)$ [(c)
and (d)] and $T_{vc}\left(\protect\omega \right)$ [(e) and (f)] ($v=a,b,c$)
as functions of the frequency of the incoming signal $\protect\omega$ for
different phase difference: (a), (c) and (e) $\protect\theta=\protect\pi/2$;
(b), (d) and (f) $\protect\theta =3\protect\pi/2$. The other parameters are $%
\Delta _{a}^{\prime }=\Delta _{b}^{\prime }=\protect\omega _{m}=10 \protect%
\gamma$, $J=G_{a}=G_{b}e^{-i\protect\theta }=\protect\gamma _{a}/2= \protect%
\gamma _{b}/2=\protect\gamma _{m}/2=\protect\gamma /2$.}
\label{fig5}
\end{figure}

\begin{figure}[tbp]
\includegraphics[bb=58 296 543 585, width=8.5 cm, clip]{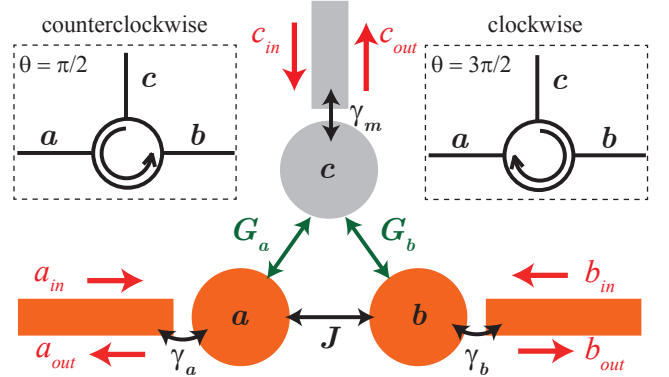}
\caption{(Color online) Schematic diagram of a three-mode optomechanical
circulator.}
\label{fig6}
\end{figure}

Using Eqs.~(\ref{Eq22})-(\ref{Eq30}), we now show the numerical results of all the scattering probabilities (nine elements) in Eqs.~(\ref{Eq21}). As shown in Fig.~\ref{fig5}, the three-mode optomechanical system shows circulator behavior: when $\theta =\pi /2$, we have $T_{ba}\left( \omega
\right) \approx T_{cb}\left( \omega \right) \approx T_{ac}\left( \omega
\right) \approx 1$ and the other scattering probabilities equal to zero at $%
\omega =\omega _{m}$ as shown in Figs.~\ref{fig5} (a), (c) and (e); when $%
\theta =3\pi /2$, we have $T_{ca}\left( \omega \right) \approx T_{ab}\left(
\omega \right) \approx T_{bc}\left( \omega \right) \approx 1$ and the other
scattering probabilities equal to zero at $\omega =\omega _{m}$ as shown in
Figs.~\ref{fig5} (b), (d) and (f). That is to say the signal is transferred
from one mode to another either counterclockwisely ($a\rightarrow
b\rightarrow c\rightarrow a$) or clockwisely ($a\rightarrow c\rightarrow
b\rightarrow a$), depending on the relative phase $\theta =\pi /2$ or $3\pi /2$ as shown in Fig.~\ref{fig6}.

The scattering matrix for the optomechanical circulator in three-mode
optomechanical systems can be obtained analytically, similar to the case for
photon and phonon circulators in Refs.~\cite{KochPRA10,HabrakenNJP12}. We
assume that $\omega _{m}\approx \Delta \gg \{J,\left\vert G_{a}\right\vert
,\left\vert G_{b}\right\vert ,\gamma _{a},\gamma _{b},\gamma _{m}$\}, then
Eqs.~(\ref{Eq8})-(\ref{Eq10}) can be simplified by rotating wave
approximation as
\begin{equation}
\frac{d}{dt}\delta a=\left( -\frac{\gamma _{a}}{2}-i\Delta _{a}^{\prime
}\right) \delta a-iG_{a}\delta c-iJ\delta b+\sqrt{\gamma _{a}}a_{\mathrm{in}%
},  \label{Eq36}
\end{equation}%
\begin{equation}
\frac{d}{dt}\delta b=\left( -\frac{\gamma _{b}}{2}-i\Delta _{b}^{\prime
}\right) \delta b-iG_{b}\delta c-iJ\delta a+\sqrt{\gamma _{b}}b_{\mathrm{in}%
},  \label{Eq37}
\end{equation}%
\begin{equation}
\frac{d}{dt}\delta c=\left( -\frac{\gamma _{m}}{2}-i\omega _{m}\right)
\delta c-iG_{a}^{\ast }\delta a-iG_{b}^{\ast }\delta b+\sqrt{\gamma _{m}}c_{%
\mathrm{in}}.  \label{Eq38}
\end{equation}%
Thus the effective Hamiltonian after linearization takes the form (effective
system is shown in Fig.~\ref{fig6})
\begin{eqnarray}
H_{\mathrm{eff}} &=&\hbar \Delta _{a}^{\prime }a^{\dag }a+\hbar \Delta
_{b}^{\prime }b^{\dag }b+\hbar \omega _{m}c^{\dag }c  \label{Eq39} \\
&&+\hbar \left( Jb^{\dag }a+|G_{a}|e^{i\theta _{a}}a^{\dag
}c+|G_{b}|e^{-i\theta _{b}}c^{\dag }b+\mathrm{H.c.}\right) .  \notag
\end{eqnarray}%
The necessary and sufficient condition of time-reversal symmetry is the
gauge-invariant phase sum is an integral multiple of $\pi $~\cite{KochPRA10}. That is,
\begin{equation}
\theta _{b}-\theta _{a}\equiv\theta=n\pi  \label{Eq40}
\end{equation}%
for real $J$, where $n$ is an integral number [also see the numerical results as shown in Figs.~\ref{fig2}(a) and \ref{fig2}(e)]. The optical nonreciprocal response and the optomechanical circulator behavior are
induced by breaking the time-reversal symmetry (i.e., $\theta _{b}-\theta
_{a}\neq n\pi $), and the optimal effect is realized at the halfway between
the time-reversal symmetric points (i.e., $\theta _{a}=0$, $\theta
_{b}=\theta =\pi /2$ or $3\pi /2$) as shown in Fig.~\ref{fig5}.
%The optomechanical circulator behavior (broken time-reversal symmetry) is realized at the halfway between the time-reversal symmetric points (i.e., $\phi _{a}+\phi _{b}=\pm \pi/2$) as shown in Fig.~\ref{fig6}.

Now we will derive the scattering matrix of the optomechanical circulator behavior analytically from the simplified linearized QLEs [Eqs.~(\ref{Eq36})-(\ref{Eq38})]. Let us transform the linearized equations into the frequency
domain,
\begin{equation}
\left( M^{\prime }-i\omega I\right) \left(
\begin{array}{c}
\widetilde{\delta a} \\
\widetilde{\delta b} \\
\widetilde{\delta c}%
\end{array}%
\right) =\left(
\begin{array}{c}
\sqrt{\gamma _{a}}\widetilde{a_{\mathrm{in}}} \\
\sqrt{\gamma _{b}}\widetilde{b_{\mathrm{in}}} \\
\sqrt{\gamma _{m}}\widetilde{c_{\mathrm{in}}}%
\end{array}%
\right) ,  \label{Eq41}
\end{equation}%
where
\begin{equation}
M^{\prime }=\left(
\begin{array}{ccc}
\frac{\gamma _{a}}{2}+i\Delta _{a}^{\prime } & iJ & iG_{a} \\
iJ & \frac{\gamma _{b}}{2}+i\Delta _{b}^{\prime } & iG_{b} \\
iG_{a}^{\ast } & iG_{b}^{\ast } & \frac{\gamma _{m}}{2}+i\omega _{m}%
\end{array}%
\right) .  \label{Eq42}
\end{equation}%
In the conditions for the optimal optomechanical circulator, i.e. $\omega=\Delta_{a}^{\prime }=\Delta _{b}^{\prime }=\omega _{m}$ and $J=G_{a}=G_{b}e^{-i\theta }=\gamma _{a}/2=\gamma _{b}/2=\gamma _{m}/2=\gamma
/2$, we have
\begin{equation}
\frac{\gamma }{2}\left(
\begin{array}{ccc}
1 & i & i \\
i & 1 & ie^{i\theta } \\
i & ie^{-i\theta } & 1%
\end{array}%
\right) \left(
\begin{array}{c}
\widetilde{\delta a} \\
\widetilde{\delta b} \\
\widetilde{\delta c}%
\end{array}%
\right) =\left(
\begin{array}{c}
\sqrt{\gamma }\widetilde{a_{\mathrm{in}}} \\
\sqrt{\gamma }\widetilde{b_{\mathrm{in}}} \\
\sqrt{\gamma }\widetilde{c_{\mathrm{in}}}%
\end{array}%
\right) .  \label{Eq43}
\end{equation}%
By choosing $\theta =\pi /2$, the scattering matrix is given through
\begin{equation}
\left(
\begin{array}{c}
\widetilde{a_{\mathrm{out}}} \\
\widetilde{b_{\mathrm{out}}} \\
\widetilde{c_{\mathrm{out}}}%
\end{array}%
\right) =\left(
\begin{array}{ccc}
0 & 0 & -i \\
-i & 0 & 0 \\
0 & -1 & 0%
\end{array}%
\right) \left(
\begin{array}{c}
\widetilde{a_{\mathrm{in}}} \\
\widetilde{b_{\mathrm{in}}} \\
\widetilde{c_{\mathrm{in}}}%
\end{array}%
\right) .  \label{Eq44}
\end{equation}%
By choosing $\theta =3\pi /2$, we can get the scattering matrix through
\begin{equation}
\left(
\begin{array}{c}
\widetilde{a_{\mathrm{out}}} \\
\widetilde{b_{\mathrm{out}}} \\
\widetilde{c_{\mathrm{out}}}%
\end{array}%
\right) =\left(
\begin{array}{ccc}
0 & -i & 0 \\
0 & 0 & -1 \\
-i & 0 & 0%
\end{array}%
\right) \left(
\begin{array}{c}
\widetilde{a_{\mathrm{in}}} \\
\widetilde{b_{\mathrm{in}}} \\
\widetilde{c_{\mathrm{in}}}%
\end{array}%
\right) .  \label{Eq45}
\end{equation}%
Equation~(\ref{Eq44}) shows clearly a perfect circulator with the signal transferring counterclockwisely ($a\rightarrow b\rightarrow c\rightarrow a$) for $\theta =\pi /2$ and Eq.~(\ref{Eq45}) also describes an ideal circulator but with the signal transferring clockwisely ($a\rightarrow c\rightarrow b\rightarrow a$) for $\theta =3\pi /2$.
These agree well with the numerical results shown in Fig.~\ref{fig5}.

\begin{figure}[tbp]
\includegraphics[bb=60 420 486 594, width=8.5 cm, clip]{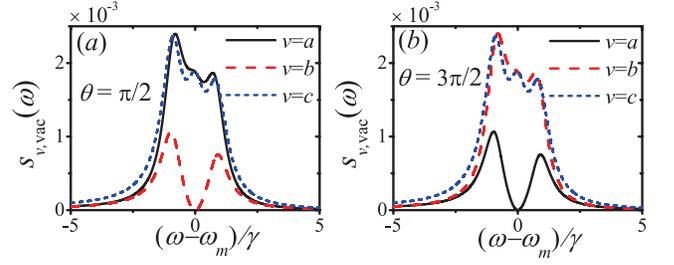}
\caption{(Color online) The vacuum noise spectrum $s_{v,\mathrm{vac}}\left(
\protect\omega \right)$ ($v=a,b,c$) as a function of the frequency of the
incoming signal $\protect\omega$ for different phase difference: (a) $%
\protect\theta= \protect\pi/2$; (b) $\protect\theta=3\protect\pi/2$. The other
parameters are $\Delta _{a}^{\prime }=\Delta _{b}^{\prime }=\protect\omega %
_{m}=10 \protect\gamma$, $J=G_{a}=G_{b}e^{-i\protect\theta }=\protect\gamma %
_{a}/2= \protect\gamma _{b}/2=\protect\gamma _{m}/2=\protect\gamma /2$.}
\label{fig7}
\end{figure}

Finally, we discuss the effects of the vacuum noise spectrum $s_{v,\mathrm{vac}}\left( \omega \right) $ given by Eqs.~(\ref{Eq31})-(\ref{Eq33}). The vacuum noise spectrum $s_{v,\mathrm{vac}}\left( \omega \right) $ ($v=a,b,c$) as a function of the frequency of the incoming signal $\omega $ is shown in Fig.~\ref{fig7}. The effects of the vacuum noises are so small that they are insignificant even for the input signals of single-photon (single-phonon) level (about $0.2\%$ at $\omega =\omega _{m}=10\gamma $). The physical origin of the vacuum noise in the output spectrum is the anti-rotating-wave interactions between the optical and the mechanical modes [included in Eq.~(\ref{Eq11})]. The suppression of the vacuum noise for $b$ mode ($a$ mode) at $\omega =\omega _{m}$ as $\theta =\pi /2$ ($\theta =3\pi /2$) is the consequence of the rotating-wave approximation for the interaction between the two optical modes.

\section{Conclusions}

In summary, we have shown the optical nonreciprocity in a three-mode
optomechanical system. We demonstrated that the nonreciprocal response is
enabled by tuning the phase difference between the optomechanical coupling rates
to induce the time-reversal symmetry breaking of the system. Then we show that the three-mode optomechanical system can also be used as a three-port
optomechanical circulator for two optical modes and one mechanical mode.
Further, we note that the three-mode optomechanical system can work in the
single-photon level and be integrated into a chip. The three-port
optomechanical circulator might eventually provide the basis for
applications on quantum information processing or quantum simulation~\cite{NunnenkampNJP11}.

\vskip 2pc \leftline{\bf Note added} In the preparation of this work, we became aware of a related paper by Metelmann and Clerk~\cite{Metelmann15arx}.

%\section{Acknowledgement}
\vskip 2pc \leftline{\bf Acknowledgement}

We thank W. H. Hu for fruitful discussions. This work is supported by the
Postdoctoral Science Foundation of China (under Grant No. 2014M550019), the
National Natural Science Foundation of China (under Grants No. 11422437, No. 11174027, and No. 11121403) and the National Basic Research Program of China (under Grants No. 2012CB922104 and No. 2014CB921403).

\bibliographystyle{apsrev}
\bibliography{ref}

\begin{thebibliography}{99}
\bibitem{PottonRPP04} R. J. Potton, Rep. Prog. Phys.~\textbf{67}, 717 (2004); I.V. Shadrivov, V. A. Fedotov, D. A. Powell, Y. S. Kivshar, and N. I. Zheludev, New J. Phys.~\textbf{13}, 033025 (2011).

\bibitem{FujitaAPL00} J. Fujita, M. Levy, R. M. Osgood, L.Wilkens, and H. D\"{o}tsch, Appl. Phys. Lett.~\textbf{76}, 2158 (2000); R. L. Espinola, T. Izuhara, M. C. Tsai, R. M. Osgood Jr., H. D\"{o}tsch, Opt. Lett.~\textbf{29}, 941 (2004); T. R. Zaman, X. Guo, R. J. Ram, Appl. Phys. Lett.~\textbf{90}, 023514 (2007); F. D. M. Haldane and S. Raghu, Phys. Rev. Lett.~\textbf{100}, 013904 (2008); Y. Shoji, T. Mizumoto, H. Yokoi, I. Hsieh, and R. M. Osgood, Appl. Phys. Lett.~\textbf{92}, 071117 (2008); Z. Wang, Y. Chong, J. D. Joannopoulos, and M. Solja\v{c}i\'{c}, Nature (London)~\textbf{461}, 772 (2009); Y. Hadad and B. Z. Steinberg, Phys. Rev. Lett.~\textbf{105}, 233904 (2010); A. B. Khanikaev, S. H. Mousavi, G. Shvets, and Y. S. Kivshar, \emph{ibid.}~\textbf{105}, 126804 (2010); L. Bi, J. Hu, P. Jiang, D. H. Kim, G. F. Dionne, L. C. Kimerling, and C. A. Ross, Nat. Photon.~\textbf{5}, 758 (2011); Y. Shoji, M. Ito, Y. Shirato, and T. Mizumoto, Opt. Express~\textbf{20}, 18440 (2012).

\bibitem{GalloAPL01} K. Gallo, G. Assanto, K. R. Parameswaran, and M. M. Fejer, Appl. Phys. Lett.~\textbf{79}, 314 (2001); S. F. Mingaleev, Y. S. Kivshar, J. Opt. Soc. Am. B~\textbf{19}, 2241 (2002); M. Solja\v{c}i\'{c}, C. Luo, J. D. Joannopoulos, S. Fan, Opt. Lett.~\textbf{28}, 637 (2003); A. Rostami, Opt. Laser Technol.~\textbf{39}, 1059 (2007); A. Alberucci and G. Assanto, Opt. Lett.~\textbf{33}, 1641 (2008); L. Fan, J. Wang, L. T. Varghese, H. Shen, B. Niu, Y. Xuan, A. M. Weiner, and M. Qi, Science~\textbf{335}, 447 (2012); L. Fan, L. T. Varghese, J. Wang, Y. Xuan, A. M. Weiner, and M. Qi, Opt. Lett.~\textbf{38}, 1259 (2013); B. Anand, R. Podila, K. Lingam, S. R. Krishnan, S. S. S. Sai, R. Philip, and A. M. Rao, Nano Lett.~\textbf{13}, 5771 (2013).

\bibitem{BiancalanaJAP08} F. Biancalana, J. Appl. Phys.~\textbf{104}, 093113(2008); A. E. Miroshnichenko, E. Brasselet, and Y. S. Kivshar, Appl. Phys. Lett.~\textbf{96}, 063302 (2010); C. Wang, C. Zhou, and Z. Li, Opt. Express~\textbf{19}, 26948 (2011); C. Wang, X. Zhong, and Z. Li, Sci. Rep.~\textbf{2}, 674 (2012); K. Xia, M. Alamri, and M. S. Zubairy, Opt. Express~\textbf{21}, 25619 (2013); E. J. Lenferink, G. Wei, and N. P. Stern, \emph{ibid.}~\textbf{22}, 16099 (2014); Y. Yu, Y. Chen, H. Hu, W. Xue, K. Yvind, and J. Mork, arXiv:1409.3147.

\bibitem{YuNP09} Z. F. Yu and S. H. Fan, Nat. Photon.~\textbf{3}, 91 (2009); K. Fang, Z. Yu, and S. Fan, \emph{ibid.}~\textbf{6}, 782 (2012); E. Li, B. J. Eggleton, K. Fang, and S. Fan, Nat. Commun.~\textbf{5}, 3225 (2014); C. R. Doerr, N. Dupuis, and L. Zhang, Opt. Lett.~\textbf{36}, 4293 (2011); C. R. Doerr, L. Chen, and D. Vermeulen, Opt. Express~\textbf{22}, 4493 (2014); H. Lira, Z. F. Yu, S. H. Fan, and M. Lipson, Phys. Rev. Lett.~\textbf{109}, 033901 (2012); K. Fang, Z. Yu, and S. Fan, \emph{ibid.}~\textbf{108}, 153901 (2012); M. C. Munoz, A. Y. Petrov, L. O'Faolain, J. Li, T. F. Krauss, and M. Eich, \emph{ibid.}~\textbf{112}, 053904 (2014); Y. Yang, C. Galland, Y. Liu, K. Tan, R. Ding, Q. Li, K. Bergman, T. Baehr-Jones, and M. Hochberg, Opt. Express~\textbf{22}, 17409 (2014).

\bibitem{WangOE10} Q. Wang, F. Xu, Z. Y. Yu, X. S. Qian, X. K. Hu, Y. Q. Lu, and H. T. Wang, Opt. Express~\textbf{18}, 7340 (2010); M. S. Kang, A. Butsch, and P. S. J. Russell, Nat. Photon.~\textbf{5}, 549 (2011).

\bibitem{EuterNP10} C. E\"{u}ter, K. G. Makris, R. EI-Ganainy, D. N. Christodoulides, M. Segev, and D. Kip, Nat. Phys.~\textbf{6}, 192 (2010); H. Ramezani, T. Kottos, R. El-Ganainy, and D. N. Christodoulides, Phys. Rev. A~\textbf{82}, 043803 (2010); L. Feng, M. Ayache, J. Q. Huang, Y. L. Xu, M. H. Lu, Y. F. Chen, Y. Fainman, and A. Scherer, Science~\textbf{333}, 729 (2011); B. Peng, S. K. \"{O}zdemir, F. Lei, F. Monifi, M. Gianfreda, G. L. Long, S. H. Fan, F. Nori, C. M. Bender, and L. Yang, Nat. Phys.~\textbf{10}, 394 (2014); J. H. Wu, M. Artoni, and G. C. La Rocca, Phys. Rev. Lett.~\textbf{113}, 123004 (2014).

\bibitem{WangPRL13} D. W. Wang, H. T. Zhou, M. J. Guo, J. X. Zhang, J. Evers, and S. Y. Zhu, Phys. Rev. Lett.~\textbf{110}, 093901 (2013); S. A. R. Horsley, J. H. Wu, M. Artoni, and G. C. La Rocca, \emph{ibid.}~\textbf{110}, 223602 (2013).

\bibitem{ShenPRL11} Y. Shen, M. Bradford, and J. T. Shen, Phys. Rev. Lett.~\textbf{107}, 173902 (2011); K. Xia, G. Lu, G. Lin, Y. Cheng, Y. Niu, S. Gong, and J. Twamley, Phys. Rev. A~\textbf{90}, 043802 (2014); H. Z. Shen, Y. H. Zhou, and X. X. Yi, \emph{ibid.}~\textbf{90}, 023849 (2014).

\bibitem{AspelmeyerARX13} T. J. Kippenberg and K. J. Vahala, Science~\textbf{321}, 1172 (2008); F. Marquardt and S. M. Girvin, Physics~\textbf{2}, 40 (2009); M. Aspelmeyer, P. Meystre, and K. Schwab, Phys. Today~\textbf{65}, 29 (2012); M. Aspelmeyer, T. J. Kippenberg, and F. Marquardt, Rev. Mod. Phys.~\textbf{86}, 1391 (2014).

\bibitem{ManipatruniPRL09} S. Manipatruni, J. T. Robinson, and M. Lipson, Phys. Rev. Lett.~\textbf{102}, 213903 (2009).

\bibitem{HafeziOE12} M. Hafezi and P. Rabl, Opt. Express~\textbf{20}, 7672 (2012).

\bibitem{KochPRA10} J. Koch, A. A. Houck, K. L. Hur, and S. M. Girvin, Phys. Rev. A~\textbf{82}, 043811 (2010).

\bibitem{HabrakenNJP12} S. J. M. Habraken, K. Stannigel, M. D. Lukin, P. Zoller, and P Rabl, New J. Phys.~\textbf{14}, 115004 (2012).

\bibitem{LudwigPRL12} M. Ludwig, A. H. Safavi-Naeini, O. Painter, and F. Marquardt, Phys. Rev. Lett.~\textbf{109}, 063601 (2012); K. Stannigel, P. Komar, S. J. M. Habraken, S. D. Bennett, M. D. Lukin, P. Zoller, and P. Rabl, \emph{ibid.}~\textbf{109}, 013603 (2012); Y. D. Wang and A. A. Clerk, \emph{ibid.}~\textbf{108}, 153603 (2012); L. Tian, \emph{ibid.}~\textbf{108}, 153604 (2012); W. J. Gu and G. X. Li, Phys. Rev. A~\textbf{87}, 025804 (2013).

\bibitem{ThompsonNat08} J. D. Thompson, B. M. Zwickl, A. M. Jayich, F. Marquardt, S. M. Girvin, and J. G. E. Harris, Nature (London)~\textbf{452}, 72 (2008); G. Heinrich, J. G. E. Harris, and F. Marquardt, Phys. Rev. A~\textbf{81}, 011801(R) (2010); H. Z. Wu, G. Heinrich, and F. Marquardt, New J. Phys.~\textbf{15}, 123022 (2013).

\bibitem{LinPRL09} Q. Lin, J. Rosenberg, X. Jiang, K. J. Vahala, and O. Painter, Phys. Rev. Lett.~\textbf{103}, 103601 (2009); S. Weis, R. Rivi\`{e}re. S. Del\'{e}glise, E. Gavartin, O. Arcizet, A. Schliesser, and T. J. Kippenberg, Science~\textbf{330}, 1520 (2010).

\bibitem{EichenfieldNat09} M. Eichenfield, R. Camacho, J. Chan, K. J. Vahala, and O. Painter, Nature (London)~\textbf{459}, 550 (2009); M. Li, W. H. P. Pernice, and H. X. Tang, Nat. Photon.~\textbf{3}, 464 (2009); A. H. Safavi-Naeini and O. Painter, New J. Phys.~\textbf{13}, 013017 (2011); J. T. Hill, A. H. Safavi-Naeini, J. Chan, and O. Painter, Nat. Commun.~\textbf{3}, 1196 (2012).

\bibitem{TeufelNat11} J. D. Teufel, D. Li, M. S. Allman, K. Cicak, A. J. Sirois, J. D. Whittaker, and R. W. Simmonds, Nature (London)~\textbf{471}, 204 (2011); F. Massel, S. Un Cho, J.-M. Pirkkalainen, P. J. Hakonen, T. T. Heikkil\"{a}, and M. A. Sillanp\"{a}\"{a}, Nat. Commun.~\textbf{3}, 987 (2012); T. A. Palomaki, J. D. Teufel, R. W. Simmonds, and K. W. Lehnert, Science~\textbf{342}, 710 (2013); J. Suh, A. J. Weinstein, C. U. Lei, E. E. Wollman, S. K. Steinke, P. Meystre, A. A. Clerk, and K. C. Schwab, \emph{ibid.}~\textbf{344}, 1262 (2014).

\bibitem{GardinerPRA85} C. W. Gardiner and M. J. Collett, Phys. Rev. A~\textbf{31}, 3761 (1985).

\bibitem{AgarwalPRA12} G. S. Agarwal and S. Huang, Phys. Rev. A 85, 021801(R) (2012).

\bibitem{NunnenkampNJP11} A. Nunnenkamp, J. Koch, and S. M. Girvin, New J. Phys.~\textbf{13}, 095008 (2011); A. L. C. Hayward, A. M. Martin, and A. D. Greentree, Phys. Rev. Lett.~\textbf{108}, 223602 (2012); R. O. Umucal{\i}lar and I. Carusotto, \emph{ibid.}~\textbf{108}, 206809 (2012); I. M. Georgescu, S. Ashhab, and F. Nori, Rev. Mod. Phys.~\textbf{86}, 153 (2014).

\bibitem{Metelmann15arx} A. Metelmann and A. A. Clerk, arXiv:1502.07274v1.
    
\end{thebibliography}

\end{document}